\documentclass[preprint,showpacs,preprintnumbers,amsmath,amssymb]{revtex4}

\usepackage{graphicx}% Include figure files
\usepackage{dcolumn}% Align table columns on decimal point
\usepackage{bm}% bold math
\usepackage{color}
\usepackage{sidecap}

\def\vq{{\bf q}}

\def\vk{{\bf k}}

\def\vr{{\bf r}}

\def\eps{\epsilon}

\newcommand{\eq}[1]{Eq.~(\ref{#1})}
\newcommand{\fig}[1]{Fig.~\ref{#1}}
\newcommand{\be}{\begin{equation}}
\newcommand{\ee}{\end{equation}}
\newcommand{\bea}{\begin{eqnarray}}

\newcommand{\eea}{\end{eqnarray}}

\begin{document}   
\title{Electronic nematic phase transition in the presence of anisotropy} 
\author{Hiroyuki Yamase}
\affiliation{Max-Planck-Institut f\"ur Festk\"orperforschung,
             Heisenbergstrasse 1, D-70569 Stuttgart, Germany}
\affiliation{National Institute for Materials Science, Tsukuba 305-0047, Japan}

\date{19 January, 2014}

\begin{abstract}
We  study the phase diagram of electronic nematic instability 
in the presence of $xy$ anisotropy. 
While a second order transition cannot occur in this case, 
mean-field theory predicts that  a first order transition occurs near van Hove filling and 
its phase boundary forms a wing structure, which we term a Griffiths wing, referring 
to his original work of He$^3$-He$^4$ mixtures. 
When crossing the wing, the anisotropy of the 
electronic system exhibits a discontinuous change, leading to 
a meta-nematic transition, i.e., the analog to a meta-magnetic 
transition in a magnetic system. 
The upper edge of the wing corresponds to a critical end line. 
It shows a non-monotonic temperature dependence as a function of 
the external anisotropy and vanishes at a quantum critical end point 
for a strong anisotropy. The mean-field phase diagram is, however, 
very sensitive to fluctuations of the nematic order parameter, yielding  
a topologically different phase diagram. 
The Griffiths wing is broken into two pieces. 
A tiny wing appears close to zero anisotropy and the other 
is realized for a strong anisotropy. Consequently three quantum 
critical end points are realized. 
We discuss that these results can be related to various materials 
including a cold atom system. 
\end{abstract}

\pacs{05.30.Fk, 71.10. Hf, 71.18.+y, 71.27.+a} 
%05.30.Fk Fermion systems and electron gas 
%71.10.Hf Non-Fermi liquid ground state, electron phase diagram and phase 
%transitions in model systems
%71.18.+y Fermi surface: calculations and measurements 
%71.27.+a Strongly correlated electron systems; heavy fermions 

%05.10.Cc Renormalization group methods
%05.30.Fk Fermion systems and electron gas 
%05.30.Rt Quantum phase transitions 
%64.40.-i General studies of phase transitions 
%64.70.-p Specific phase transitions
%64.70.Tg quantum phase transitions
%71.10.-w Theories and models of many-electron systems
%71.10 Fd Lattice fermion model
%71.10.Hf Non-Fermi liquid ground state, electron phase diagram and phase 
%transitions in model systems
%71.18.+y Fermi surface: calculations and measurements 
%71.27.+a Strongly correlated electron systems; heavy fermions 
%74.40.Kb Quantum critical phenomena 
% 74.70.Pq Ruthenates
%74.72.-h Curpate superconductors 
%75.40.-s Critical-point effects, specific heats, short-range order
%75.40.Cx Static properties (order parameter, static susceptibility, heat capacities, critical exponents) 
%74.20.Mn Nonconventional mechanism 
%74.72.Gh Hole-doped cuprates
%74.72.Ek electron-doped cuprates
%74.72.Kf Pseudogap regime 
%75.25.Dk Orbital, charge, and other orders, including coupling of these orders

\maketitle
\section{introduction}
Nematic liquid crystals are well known. Rodlike molecules 
flow like a liquid, but are always oriented to a certain direction 
in the nematic phase. This state is characterized by breaking of 
the orientational symmetry, retaining the other symmetries of the system.  
Electrons are point particles, not molecules. Nevertheless  
electronic analogs of the nematic liquid crystals were observed 
in a number of interacting electron systems: 
Two-dimensional electron gases \cite{lilly99,du99}, 
high-temperature superconductors of cuprates  \cite{kivelson03,vojta09} 
and pnictides \cite{fisher11},   
the bilayer strontium ruthenate Sr$_3$Ru$_2$O$_7$ \cite{mackenzie12}, and 
an actinide material URu$_2$Si$_2$ \cite{okazaki11}. 

The electronic nematic order couples directly to an external anisotropy, 
which is thus expected to play a crucial role in a system exhibiting nematicity. 
The external anisotropy can be controlled by applying a uniaxial  pressure, (epitaxial) strain, 
and sometimes by a crystal structure due to orthorhombicity. 
While it is generally not easy to quantify how much anisotropy is imposed on a sample, 
the anisotropy was calibrated recently by exploiting 
the piezoelectric effect \cite{chu12}. A nematic susceptibility was then extracted and 
its divergence was demonstrated near a nematic critical point. 

Motivated by the experimental progress to control the external anisotropy, 
we study a role of the external anisotropy for the electronic nematic instability. 
This fundamental issue has not been well addressed even in mean-field theory. 
In particular, we focus on the nematicity associated with a 
$d$-wave Pomeranchuk instability ($d$PI) \cite{yamase00,metzner00}, 
which is expected to exhibit interesting physics. 
In a mean-field theory in the absence of anisotropy \cite{khavkine04,yamase05}, 
the $d$PI occurs around van Hove filling with a dome-shaped transition line. 
The transition is of second order at high temperatures  
and changes to first order at low temperatures. 
The end points of the second order line are tricritical points. 

The presence of a tricritical point (TCP) implies a wing structure when a conjugate field to 
the corresponding order parameter is applied to the system. 
This insight originates from the study of He$^{3}$-He$^{4}$ mixtures 
by Griffiths \cite{griffiths70}.  
However, the conjugate field to the superfluid order parameter is not accessible 
in experiments. The wing structure predicted by Griffiths, which we term the Griffiths wing, 
was not tested for He$^{3}$-He$^{4}$ mixtures. 

It was found that itinerant ferromagnetism occurs generally 
via a first order transition at low temperatures and a second order one at high temperatures 
\cite{belitz05}. The end point of the second order line is a TCP. 
The order parameter is magnetization and its conjugate field is a magnetic field 
in that case. 
Similar to Griffiths's work \cite{griffiths70}, a wing structure emerges from the first 
order transition line and extends to the side of a finite magnetic field. 
When crossing the wing, the system exhibits a jump of the magnetization, 
leading to a metamagnetic transition. 
Recently, the Griffiths wings were clearly 
observed in ferromagnetic metals such as UGe$_{2}$ \cite{kotegawa11} and UCoAl \cite{aoki11}. 

In this paper, we study Griffiths wings of an electronic nematic phase transition 
associated with the $d$PI. 
A conjugate field to the nematic order parameter is $xy$ anisotropy, which is  
accessible in experiments. By applying the anisotropy, 
we obtain a wing structure. 
However, in contrast to previous studies \cite{griffiths70,belitz05}, 
the Griffiths wing exhibits a non-monotonic temperature dependence. 
Furthermore we find that the wing structure is very sensitive to fluctuations of 
the order parameter, leading to a phase diagram 
topologically different from the mean-field result. 
These results can be related to various materials including a cold atom system. 

\section{model} 
We study electronic nematicity associated with the $d$PI in the 
presence of $xy$ anisotropy. Our minimal model reads 
\be
H=\sum_{\vk, \sigma}(\eps^{0}_{\vk} -\mu) c^{\dagger}_{\vk \sigma}c_{\vk \sigma}-
\frac{1}{2N} \sum_{\vq} g(\vq) n_{d}(\vq) n_{d}(-\vq) -\mu_{d}n_{d}({\bf 0}) \,,
\label{model}
\ee
where $c^{\dagger}_{\vk \sigma}$ ($c_{\vk \sigma}$) is the creation (annihilation) operator 
of electrons with momentum $\vk$ and spin $\sigma$, 
$\mu$ is the chemical potential, and 
$N$ is the number of sites. The kinetic energy $\eps_{\vk}^{0}$ is given by 
a usual tight binding dispersion on a square lattice, 
\be
\eps^{0}_{\vk} = -2 t (\cos k_{x}+\cos k_{y}) - 4t'\cos k_{x}\cos k_{y}\,.
\label{bareband}
\ee
The interaction term describes a $d$-wave weighted density-density interaction \cite{metzner03}; 
$n_{d}(\vq)=\sum_{\vk, \sigma}d_{\vk} c^{\dagger}_{\vk - \frac{\vq}{2} \sigma}
c_{\vk+\frac{\vq}{2}  \sigma}$ 
with a $d$-wave form factor such as $d_{\vk}=\cos k_{x}-\cos k_{y}$.  The coupling strength $g(\vq)$ 
has a peak at $\vq=0$, that is, forward scattering dominates. 
This interaction drives a $d$PI at low temperatures and is 
obtained in microscopic models such as 
the $t$-$J$ \cite{yamase00}, Hubbard \cite{metzner00,valenzuela01}, 
and general models with central forces \cite{quintanilla08}, and also from 
dipole-dipole interaction \cite{clin10}. 
A new aspect of the present study lies in the third term in \eq{model}. 
This term is expressed as 
$-\mu_{d} \sum_{\vk, \sigma} (\cos k_x - \cos k_y) c^{\dagger}_{\vk \sigma}c_{\vk \sigma}$,   
and imposes an anisotropy of 
the nearest neighbor hopping integral $t$ between the $x$ and $y$ direction 
as easily seen from Eqs.~(\ref{model}) and (\ref{bareband}). 
A value of $\mu_d$ is controlled by applying a uniaxial pressure and a strain, 
and also by an orthorhombic crystal structure. 
$\mu_{d}$ may be interpreted as the 
$d$-wave chemical potential in the sense that 
it couples to the $d$-wave weighted charge density. 
Since the order parameter of the $d$PI 
is proportional to $n_{d}({\bf 0})$, 
$\mu_{d}$ is a conjugate field to that and plays an essential role to generate a Griffiths wing 
associated with the $d$PI. 

The interaction term in \eq{model} is generated by spin-exchange \cite{yamase00,miyanaga06,edegger06,bejas12}, 
Coulomb interaction \cite{metzner00,valenzuela01,wegner02,okamoto10,su11}, 
central forces \cite{quintanilla08}, 
and dipole-dipole interaction \cite{lin10}. 
Hence various models can exhibit a strong tendency toward the $d$PI at low-energy scale, 
especially when the Fermi surface is close to saddle points around $(\pi,0)$ and $(0,\pi)$ 
where the $d$-wave form factor is enhanced. 
Our Hamiltonian (\ref{model}) is applicable to such a situation and 
is regarded as a low-energy effective model of the $d$PI, independent of 
microscopic details.  There can occur a competition with 
other tendencies such as superconductivity and magnetism in microscopic models, 
but Hamiltonian (\ref{model}) does not contain interactions other than the $d$PI. 
Thus a competing physics is beyond the scope of the present study. 
Instead we wish to clarify a role of anisotropy 
in a rather general setup, focusing on the nematic physics.  
Although the interaction term might gain 
an anisotropic term especially for a large value of $\mu_d$, we believe that the conceptional  basis 
of the Griffiths wings associated with nematicity is captured by Hamiltonian (\ref{model}). 

Hamiltonian (\ref{model}) with $\mu_d=0$, namely without anisotropy, 
was already studied in mean-field theory \cite{khavkine04,yamase05}.  
It was found \cite{yamase05} that the mean-field phase diagram of the $d$PI 
is characterized by a single energy scale. As a result, there exist various universal ratios 
charactering the phase diagram, which nicely captures experimental observations 
in Sr$_{3}$Ru$_{2}$O$_{7}$ \cite{yamase07c,yamase13}. 
The presence of momentum transfer $\vq$ in the second term in \eq{model} allows fluctuations 
around the mean-field solution. In an isotropic case ($\mu_{d}=0$), 
it was shown that nematic order-parameter fluctuations change a first order transition obtained in 
a mean-field theory into a continuous one when the fluctuations 
become sufficiently strong \cite{jakubczyk09}; 
further stronger fluctuations can even destroy completely 
the nematic instability \cite{yamase11a}. 
Nematic fluctuations close to a nematic quantum critical point lead to 
non-Fermi liquid behavior \cite{oganesyan01,metzner03,garst10}. 
It was also found that thermal nematic fluctuations near a nematic phase transition 
lead to a pronounced broadening 
of the quasi-particle peak with a strong momentum dependence 
characterized by the form factor $d_{\vk}^2$, 
leading to a Fermi-arc like feature \cite{yamase12}. 
A role of $xy$ anisotropy $(\mu_d\neq 0)$ 
was studied in the context of competition of a nematic instability and  $d$-wave pairing 
instability \cite{yamase00,okamoto10,su11}. 
It was emphasized that a given anisotropy to the system 
can be strongly enhanced due to nematic correlations. 
This feature was discussed to explain the strong anisotropy of 
magnetic excitation spectra \cite{hinkov04,hinkov07,hinkov08,yamase06,yamase09} and 
the Nernst coefficient  \cite{daou10,hackl09} in high-temperature superconductors. 
Except for these studies, a role of $xy$ anisotropy in the nematic physics is poorly understood 
even in mean-field theory. 
This issue is addressed in terms of our Hamiltonian (\ref{model}) 
including the effect of fluctuations on mean-field results.

We consider a phase diagram in the three-dimensional space spanned 
by $\mu$, $\mu_d$, and temperature $T$. The phase diagram is symmetric with respect to 
the axis of $\mu_d=0$ and is almost symmetric with respect to 
the axis of $\mu=0$ as long as $t'$ is small. 
Hence we focus on the region of $\mu>0$ and $\mu_d>0$ 
by taking $t'=0$. Considering previous studies in He$^{3}$-He$^{4}$ mixtures \cite{griffiths70} 
and ferromagnetic systems \cite{belitz05}, 
we may expect a wing structure emerging from a first order transition of the $d$PI 
by applying the field $\mu_d$.  
The upper edge of the wing  is a critical end line (CEL), 
which is determined by the condition  
\be
\frac{\partial \omega}{\partial \phi}=\frac{\partial^{2} \omega}{\partial \phi^{2}}=\frac{\partial^{3} \omega}{\partial \phi^{3}}=0 \,,
\label{CELcondition}
\ee
where $\omega$ is the Gibbs free energy per lattice sites and $\phi$ is the order parameter of the $d$PI.
Below all quantities of dimension of energy are presented in units of $t$.

\section{Mean-field analysis} 
We first study Hamiltonian (\ref{model}) in a mean-field theory. The interaction term 
is decoupled by introducing the order parameter 
\be
\phi = g n_{d}({\bf 0})\,,
\label{phi}
\ee
where $g=g({\bf 0})>0$. The mean-field Hamiltonian then reads 
\be
H_{MF}=\sum_{\vk,\sigma} \xi_{\vk}c_{\vk \sigma}^{\dagger}c_{\vk \sigma} 
+\frac{N}{2g}\phi^{2}\,,
\ee   
with the renormalized band 
\be
\xi_{\vk}=\eps_{\vk}^{0}-\mu-(\phi+\mu_{d}) d_{\vk}\,.
\label{MFband}
\ee
Obviously the conjugate field $\mu_d$ breaks $xy$ symmetry and plays 
the same role as the order parameter $\phi$. 
It is straightforward to obtain the free energy 
\be
\omega(\phi)=-\frac{2T}{N}\sum_{\vk}\log \left(1+{\rm e}^{-\xi_{\vk}/T} \right)+\frac{1}{2g}\phi^{2}\, 
\label{MFomega}
\ee
and we solve \eq{CELcondition} numerically.

 %%%%%%%%%%%%% FIG. 1 %%%%%%%%%%%%%%%%
\begin{figure}  
\includegraphics[angle=0,width=12cm]{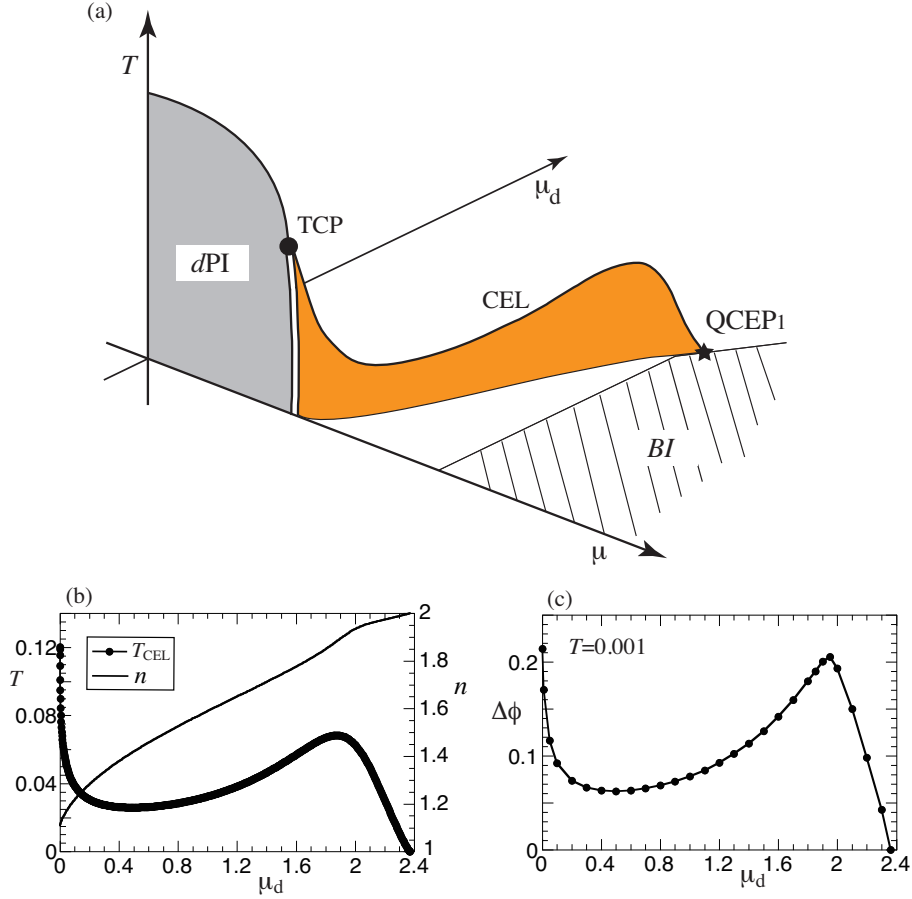} 
\caption{\label{MFresults} 
(Color online) Mean-field results.      
(a) Schematic phase diagram in the space spanned by $\mu$, $\mu_d$, and $T$. 
At $\mu_d=0$ a $d$PI occurs around van Hove filling, from which $\mu$ is measured. 
The transition is of second order at high $T$ (solid line) and of first order at low $T$ (double line). 
The solid circle denotes the TCP. The BI state is realized in the striped region. 
The wing (colored in orange) stands almost vertically 
on the plane of $\mu$ and $\mu_d$ close to van Hove filling and 
vanishes at the QCEP$_{1}$; the index 1 implies that the system is almost one dimensional. 
The upper edge of the wing (solid line) is a CEL. 
(b) Temperature of the CEL ($T_{\rm CEL}$) as a function of $\mu_d$; 
The wing is projected on the plane of $\mu_d$ and $T$. 
The electron density at $T_{\rm CEL}$ is also plotted. 
(c) Jump of the nematic order parameter across the wing at $T=0.001$. 
}
\end{figure}
%%%%%%%%%%%%%%%%%%%%%%%%%%%%%%%%%

Figure~\ref{MFresults}(a) is a schematic mean-field phase diagram \cite{misc-fig1}. 
At zero anisotropy ($\mu_d=0$) a $d$PI occurs via a first 
order transition at low $T$ as already found in previous 
studies \cite{khavkine04,yamase05}.  With increasing $\mu$, the band is eventually filled up and 
the band insulating (BI) state is realized in the striped region. 
Its phase boundary is given by $\mu=4 t$ for $\mu_{d}<2$ and $\mu=2\mu_{d}$  for 
$\mu_{d}>2$ \cite{misc-BI}. 
A wing emerges from the first order line and extends to a region of a finite $\mu_{d}$. 
The wing stands nearly vertically on the plane of $\mu_{d}$ and $\mu$ plane, and 
evolves close to van Hove filling on that plane. 
To see the wing structure more closely, we project the CEL on the plane of $\mu_d$ and $T$ 
in \fig{MFresults}(b). The temperature of the CEL, $T_{\rm CEL}$, 
is rapidly suppressed by applying the anisotropy $\mu_d$, but does not go to zero. 
It recovers to form a broad peak around $\mu_d=2$  and eventually vanishes when it touches 
the BI phase, leading to a quantum critical end point (QCEP) there. 
In fact, the electron density  becomes two at the QCEP as seen in \fig{MFresults}(b). 
When the system crosses the wing, the nematic order parameter exhibits a jump, 
leading to a meta-nematic transition. 
Such a jump, $\Delta \phi$, is plotted in \fig{MFresults}(c)  
along the bottom of the wing as a function of $\mu_d$. 
The magnitude of the jump 
exhibits behavior similar  to $T_{\rm CEL}$. 
It is interesting that $\Delta \phi$ around $\mu_{d}=2$ 
becomes comparable to that at $\mu_d=0$ in spite of the presence of 
a large external anisotropy. 

Figure~\ref{MFresults} can be understood in terms of the $d$-wave weighted density of states, 
$N_{d}(\mu)=\frac{1}{N}\sum_{\vk} d_{\vk}^{2} \delta(\xi_{\vk})$. 
This quantity appears in the second condition in \eq{CELcondition}, i.e., $\frac{\partial^{2} \omega}{\partial \phi^{2}}=0$, and diverges at van Hove filling unless the $d$-wave form factor vanishes 
at the saddle points. 
One can easily check that \eq{CELcondition} is fulfilled 
close to such van Hove filling, leading to the Griffiths wing there. 
While the field $\mu_{d}$ modifies a band structure 
as $t_x=t(1+\mu_d/2t)$ and $t_y=t(1-\mu_d/2t)$, 
the saddle points of the non-interacting band dispersion remain at 
$(\pi,0)$ and $(0,\pi)$ as long as $\mu_d <2$. 
However, for $\mu_{d}>2$, the saddle pints shift to $(\pi,\pi)$ and $(0,0)$. 
Around $\mu_d=2$, therefore, the band becomes very flat, yielding 
a substantial enhancement of the density of states. This is the reason why $T_{\rm CEL}$ as well as 
$\Delta \phi$ exhibits a peak around $\mu_d=2$;  the peak position is slightly deviated from 
$\mu_d=2$ because of the presence of a finite order parameter $\phi$. 
Since the saddle points $(\pi,\pi)$ and $(0,0)$ do not contribute to 
$N_{d}(\mu)$ because the $d$-wave form factor $d_{\vk}$ vanishes there, 
$T_{\rm CEL}$ is suppressed for $\mu_{d}>2$ 
and ultimately vanishes near the band edge. 
Since $\mu_d$ is very large close to the QCEP$_{1}$,   
the system is almost one dimensional \cite{misc-largemud}. 
Therefore we find a remarkable property that the Griffiths wing 
interpolates between a two- and (effectively) one-dimensional system by controlling the anisotropy.

\section{effect of order-parameter fluctuations} 
In a mean-field theory we pick up the component with $\vq={\bf 0}$ in Hamiltonian (\ref{model}) 
[see also \eq{phi}]. 
Contributions from a finite $\vq$ describe order-parameter fluctuations 
around the mean-field results. 
We address such fluctuation effects on the mean-field phase diagram. 
Since the Griffiths wing is realized near the van Hove singularity, 
a usual polynomial expansion of the order-parameter potential \cite{hertz76} is not valid there. 
To circumvent such a problem, we employ a functional renormalization-group (fRG) 
scheme \cite{metzner12}. This scheme allows us to analyze fluctuations without any expansion 
of the potential \cite{wetterich93} and was successfully applied to studies of 
fluctuation effects of the $d$PI in an isotropic case ($\mu_d=0$) \cite{jakubczyk09, yamase11a}.  

We use a path-integral formalism and follow a usual procedure 
to derive an order-parameter action \cite{hertz76}. 
We first decouple the fermionic  interaction in \eq{model} 
by introducing a Hubbard-Stratonovich field associated with the fluctuating 
order parameter of the $d$PI and then integrate fermionic degrees of freedom. 
Because we are interested in low-energy, long-wavelength fluctuations of the $d$PI, 
we retain the leading momentum and frequency dependencies of the two-point function and 
neglect such dependencies in high-order vertex functions. 
The resulting order-parameter action becomes 
\be
S[\phi]=\frac{1}{2} \sideset{}{'}\sum_{q} \left[
\phi_{q}\left(A_0 \frac{|\omega_n|}{|\vq|} + Z_{0} \vq^2 \right) \phi_{-q} 
\right] + {\cal U}[\phi]\,, 
\label{bareaction}
\ee
where $\phi_{q}$  with $q=(\vq,\omega_n)$ denotes the momentum representation of the 
order-parameter field $\phi$ and 
$\omega_n = 2\pi n T$ with integer  $n$ denotes the bosonic Matsubara frequencies. 
The approximation scheme of our action (\ref{bareaction}) 
corresponds to the next-leading order of derivative expansion. 
Hence the momenta and frequencies contributing to the action $S[\phi]$ 
should be restricted by the cutoff $\Lambda_0$ to the
region 
$\frac{A_0|\omega_{n}|}{Z_0 |\vq|} + \vq^{2} \leq \Lambda_0^2$, 
as emphasized by adding the prime in the summation 
in \eq{bareaction}.  
In the fermionic representation, $\Lambda_0$ may be related to the maximal 
momentum transfer allowed by the interaction in the second term in 
Hamiltonian (\ref{model}). 
If $\Lambda_{0}$ is set to be zero, the action (\ref{bareaction}) 
reproduces the mean-field theory. 
Physically, therefore, the value of $\Lambda_{0}$ 
controls the strength of order-parameter fluctuations.  The effective potential ${\cal U}[\phi]$ 
is given by ${\cal U}[\phi] = \int_0^{\frac{1}{T}} d\tau \int d^2 \vr \, U[\phi(\vr,\tau)]$,  
where $U(\phi)$ is equal to the mean-field potential [\eq{MFomega}] 
and we do not expand it in powers of $\phi$, in contrast to the usual case \cite{hertz76}. 

We carry out calculations in the one-particle irreducible 
scheme of the fRG by computing the flow
of the effective action $\Gamma^{\Lambda}[\phi]$ 
in the presence of an infrared cutoff $\Lambda$ \cite{metzner12}; 
$\Lambda$ is a quantity independent of $\Lambda_{0}$. 
In this scheme $\Gamma^{\Lambda}[\phi]$ interpolates between the bare action 
$S[\phi]$ [\eq{bareaction}] at the ultraviolet cutoff $\Lambda^{\rm UV}$ and the 
thermodynamic potential $\omega(\phi)$ in the limit of $\Lambda \to 0$. 
Its evolution is given by the functional exact flow equation \cite{wetterich93}, 
\be
\partial_{\Lambda} \Gamma^{\Lambda}\left[\phi\right]=
\frac{1}{2}{\rm tr}\frac{\partial_{\Lambda}R^{\Lambda}}{\Gamma_2^{\Lambda} \left[\phi\right] + R^{\Lambda}}\,,
 \label{wetterich} 
\ee
where $R^{\Lambda}$ is a regulator 
and $\Gamma_2^{\Lambda}\left[\phi\right] = 
 \delta^2\Gamma^{\Lambda}[\phi]/\delta \phi^2$. 
We choose a Litim-type regulator \cite{litim01} used in previous works \cite{jakubczyk09,yamase11a}:  
%\be
$R^{\Lambda}= \left[Z\left(\Lambda^{2} -\vq^2\right)-A \frac{|\omega_{n}|}{|\vq|} \right]
\theta \left(Z\left(\Lambda^{2} -\vq^2\right)-A \frac{|\omega_{n}|}{|\vq|}\right)$.  
%\ee 
A form of $\Gamma^{\Lambda}[\phi]$ is highly complicated and we parameterize it as 
\be
\Gamma^{\Lambda}[\phi]= \frac{1}{2} \sideset{}{'}\sum_{q} \left[
\phi_{q}\left(A^{\Lambda} \frac{|\omega_n|}{|\vq|} + Z^{\Lambda} \vq^2 \right) \phi_{-q} 
\right] + {\cal U}^{\Lambda}[\phi]\,, 
\label{action}
\ee
the same functional form as \eq{bareaction}. 
We allow flows of $Z^{\Lambda}$ and ${\cal U}^{\Lambda}$, but discard 
the flow of $A^{\Lambda}$, which is of minor importance \cite{jakubczyk08} 
and is assumed to be $A^{\Lambda}=A_{0}$. 
Inserting \eq{action} into \eq{wetterich} and evaluating the resulting equations for 
uniform fields, we obtain the flow equations 
of $Z^{\Lambda}$ and $U^{\Lambda}$; their derivations 
contain technical details and thus are left to the Appendix. 
By solving these flow equations numerically by reducing $\Lambda$ from 
$\Lambda^{\rm UV}$ to zero, we can determine the thermodynamic potential 
$\omega(\phi)$, which carries over the effect of nematic order-parameter fluctuations. 
To determine the Griffiths wing, we then 
search for a solution of \eq{CELcondition} in the three-dimensional space spanned 
by $\mu$, $\mu_d$, and $T$. 
All these computations are performed numerically and 
require highly accurate numerics, otherwise 
higher order derivatives of the free energy, which are contained in \eq{CELcondition}, 
would not become smooth enough to conclude a phase diagram.

%%%%%%%%%%%%% FIG. 2 %%%%%%%%%%%%%%%%%
\begin{figure}[t]  
\includegraphics[angle=0,width=12cm]{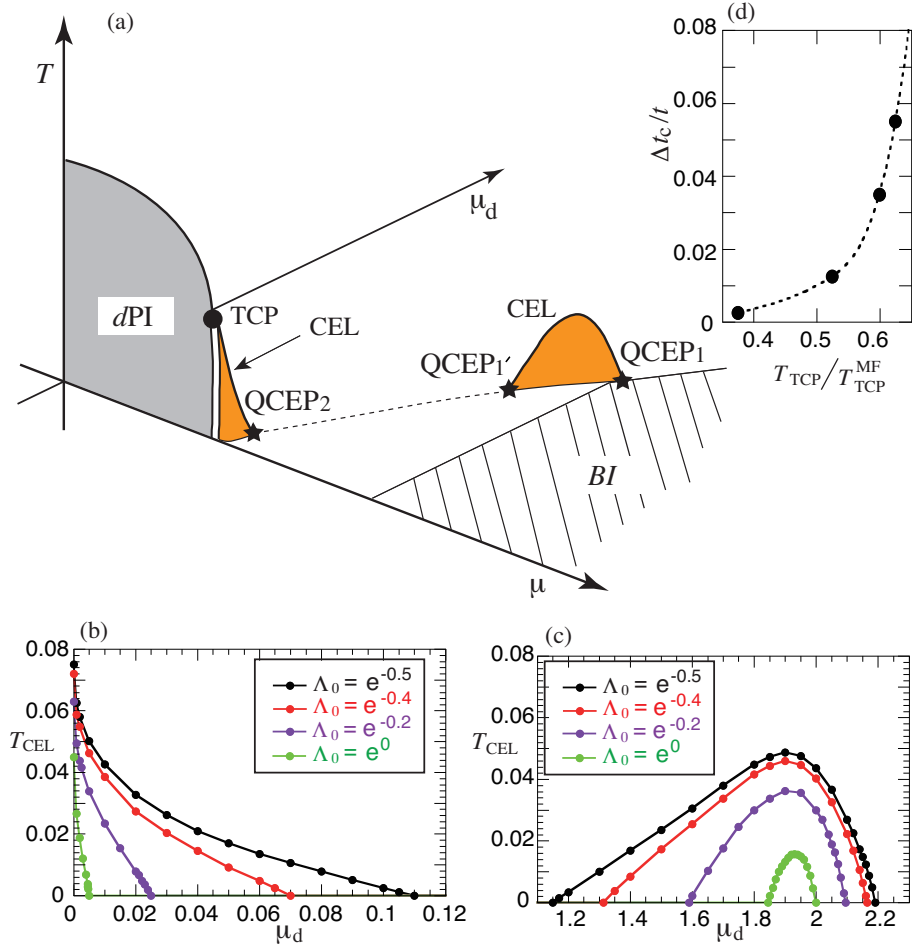} 
\caption{\label{fRGresults} 
(Color online) Results in the presence of weak nematic order-parameter fluctuations.   
(a) Schematic  phase diagram. The $d$PI phase 
is slightly suppressed by fluctuations. 
The wing obtained in \fig{MFresults}(a)   
is broken into two pieces: one tiny wing 
close to $\mu_d=0$ and the other wing for a large $\mu_{d}$. 
The indices of QCEP$_{2}$ and QCEP$_{1'}$ imply that the system can be regarded 
to be two and quasi-one dimensional, respectively, at the QCEP. 
The dashed line corresponds to a crossover. 
The BI phase is assumed to be the same as the mean-field result. 
(b) and (c) The wings are projected on the plane of $\mu_d$ and $T$ 
for several choices of the cutoff $\Lambda_0$. 
(d) The critical anisotropy of the hopping integral 
to obtain the QCEP$_{2}$ 
as a function of the ratio of the tricritical temperature and its mean-field value. 
}
\end{figure}
%%%%%%%%%%%%%%%%%%%%%%%%%%%%%%%%%%%

Figure~\ref{fRGresults}(a)  is a schematic phase diagram in the presence of 
order-parameter fluctuations. 
The $d$PI phase diagram at zero anisotropy 
is slightly suppressed by fluctuations, 
but retains essentially the same features as the mean-field result.  
Applying the anisotropy $\mu_d$, the CEL is rapidly suppressed, 
leading to a tiny wing terminating at a QCEP$_{2}$. 
We then  have a crossover region depicted by the dashed line. 
The order parameter of the $d$PI shows a rapid change, but without a jump, 
by crossing the dashed line. With further increasing $\mu_d$,  
another wing emerges with two QCEPs. 
While the Griffiths wing might seem to be broken up into two separate pieces by 
fluctuations, the two wings are actually connected via a crossover line [dashed line in \fig{fRGresults}(a)] 
as reminiscence of a single Griffiths wing in the absence of fluctuations

These results may be understood as originating from a unique feature of the mean-field result, 
namely the suppression of $T_{\rm CEL}$ 
in an intermediate region of $\mu_{d}$ in \fig{MFresults}(b). 
Given that $\phi$ enters the renormalized dispersion \eq{MFband}, 
it is easily expected that fluctuations of $\phi$ 
blur the van Hove singularity and yield the suppression of the density of states. 
Consequently, relatively low $T_{\rm CEL}$ obtained in the mean-field theory is 
easily suppressed to become zero, leading to the breaking of the Griffiths wing. 

In Figs.~\ref{fRGresults}(b) and (c) two Griffiths wings are projected 
on the plane of $\mu_d$ and $T$. The results are shown for several choices of 
the cutoff $\Lambda_{0}$, which 
controls the strength of fluctuations; 
a larger $\Lambda_{0}$ means stronger fluctuations.  
When $\Lambda_0$ becomes larger, 
the CELs are suppressed more as expected. 
This suppression is, however,  quite remarkable. 
To quantify the suppression, we consider the critical 
external anisotropy, $\Delta t_{c} /t =\frac{t_x-t_y}{t_x+t_y}=\mu_{d}/2t$, 
to obtain a QCEP. 
We plot $\Delta t_{c} /t$ in 
\fig{fRGresults}(d) as a function of 
the ratio of the tricritical temperature and 
its mean-field value ($T_{\rm TCP}/T_{\rm TCP}^{\rm MF}$) 
as the strength of fluctuations. 
We see that when $T_{\rm TCP}$ is suppressed by fluctuations, 
for example, by half, 
a very small anisotropy ($\Delta t_{c}/t \approx 0.01)$ is sufficient to yield a QCEP.  
Furthermore the strength of fluctuations to realize the QCEP is weak 
in the sense that the $d$PI phase diagram at $\mu_d=0$ is still 
well captured by mean-field theory. 
Therefore we conclude that the Griffiths wing is very sensitive to 
fluctuations and the QCEP$_{2}$ can be reached with a weak anisotropy 
even though the transition is of first order at zero anisotropy.  
This is sharply different from the mean-field result [\fig{MFresults}(a)] where 
a QCEP can be reached with a very strong anisotropy, i.e., 
$\Delta t_{c}/t \approx 1.2$ for $T_{\rm TCP}/T_{\rm TCP}^{\rm MF}=1$.

\section{Discussions}
As mentioned in Sec.~II, our Hamiltonian (\ref{model}) is a low-energy effective model 
of an electronic nematic phase transition in the presence of $xy$ anisotropy.  
It addresses a situation where a nematic tendency becomes dominant at low energy, 
independent of microscopic details. 
Usually mean-field theory is powerful to discuss actual materials at least about 
qualitative features. However, the Griffiths wing turns out to be 
sensitive even to weak fluctuations, leading to the phase diagram (\fig{fRGresults}) 
qualitatively different from the mean-field phase diagram (\fig{MFresults}). 
Since we may always have at least weak fluctuations of the order parameter in 
actual materials, \fig{fRGresults} is expected to be more realistic than \fig{MFresults}. 
Hence we bear \fig{fRGresults} in mind 
and discuss relevance to various systems 
as well as theoretical implications for future studies.

{\it Cuprates.} 
Neutron scattering experiments showed 
that the magnetic excitation spectrum becomes anisotropic 
in momentum space.  
The anisotropy observed in 
YBa$_{2}$Cu$_{3}$O$_{6.85}$ and YBa$_{2}$Cu$_{3}$O$_{6.6}$ \cite{hinkov04,hinkov07} 
is relatively weak and is well captured in terms of competition of the tendency 
toward the $d$PI and pairing correlations \cite{yamase06}. 
For YBa$_{2}$Cu$_{3}$O$_{6.45}$, however, Ref.~\onlinecite{hinkov08} reported 
a very strong anisotropy, 
which could not be interpreted in the same theory as Ref.~\onlinecite{yamase06}. 
Instead two different theories were proposed: 
one invoking the presence of a nematic quantum critical point \cite{kim08} and the other  
invoking a dominant nematic tendency over the pairing tendency \cite{yamase09}. 
The point is that the observed anisotropy seems to suddenly change  
by crossing the oxygen concentration around $6.45$. 

Microscopic models of cuprates such as 
$t$-$J$ and Hubbard exhibit the $d$PI tendency as shown by various approximation schemes: 
slave-boson mean-field theory \cite{yamase00}, exact diagonalization \cite{miyanaga06}, 
variational Monte Carlo \cite{edegger06}, 
dynamical mean-field theory \cite{okamoto10},  dynamical cluster approximation \cite{su11}, 
and a large-$N$ expansion \cite{bejas12}. 
Furthermore our low-energy effective interaction [the second term in \eq{model}] can be 
obtained from those microscopic models \cite{yamase00,metzner00,valenzuela01}. 
The $t$-$J$ model \cite{yamase00,edegger06,bejas12} 
actually exhibits a nematic tendency very  similar to the present mean-field result [\fig{MFresults} for $\mu_d=0$]. 
The nematicity in the $t$-$J$ model is strongly enhanced by approaching half-filling, which 
corresponds to van Hove filling of the spinon dispersion in the slave-boson mean-field theory \cite{lee06}.  
While a competition with other tendencies is beyond the scope of the present theory, 
our low-energy theory is expected to capture the essential feature at least 
associated with the nematicity.

Superconducting samples of Y-based cuprates have an intrinsic $xy$ anisotropy 
coming from the CuO chain structure and its anisotropy is estimated around 
$\mu_{d} =0.03$-$0.04$ \cite{yamase06}. Hence Y-based cuprates are 
located along the axis of a small  $\mu_d$ in \fig{fRGresults}(a).  
With decreasing $\mu$ (hole picture), 
namely decreasing the oxygen concentration, the system can cross the tiny wing 
or pass close to the QCEP$_{2}$, 
which may explain a sudden change of the anisotropy 
observed in the magnetic excitation spectrum \cite{hinkov08}. 
In this context, it is interesting to explore a possibility that the presence of 
a nematic quantum critical point assumed in previous studies \cite{kim08,huh08} 
can be associated with the QCEP$_{2}$. 

{\it Ruthenates.} 
The strontium ruthenate Sr$_3$Ru$_2$O$_7$ 
exhibits an electronic nematic instability around a magnetic field 8 T \cite{mackenzie12}. 
The system is tetragonal, namely $\mu_{d}=0$. 
The band structure calculations show six Fermi surfaces at zero magnetic field \cite{hase97,singh01}. 
There is a  two-dimensional Fermi surface very close to the momenta $(\pi,0)$ and $(0,\pi)$, 
which contribute to the large density of states near the Fermi energy. Hence focusing on such 
a two-dimensional band near van Hove filling, the nematic instability in Sr$_3$Ru$_2$O$_7$ is frequently 
discussed in terms of a one-band model. The interaction term in 
Hamiltonian (\ref{model})  is employed in various theoretical studies 
for Sr$_{3}$Ru$_{2}$O$_{7}$ \cite{kee05,doh07,puetter07,yamase09ab,ho08,fischer10}, 
which indeed capture major aspects of the experimental phase diagram 
except for a slope of the first order transition \cite{yamase07b,yamase07c}. 
Although the Zeeman field is not considered in the present theory, it simply 
splits the Fermi surfaces of the spin-up and spin-down band 
and then tunes the Fermi surface of either spin 
closer to van Hove filling, a very similar role to the chemical potential. 
In fact, explicit calculations including the Zeeman field 
confirm this consideration \cite{kee05,yamase07b,yamase07c,yamase13}. 
Therefore, on the basis of the present study (\fig{fRGresults}), we predict an emergence of 
a tiny Griffiths wing close to $\mu_{d}=0$ by applying 
a strain along the $x$ or $y$ direction in Sr$_3$Ru$_2$O$_7$ \cite{misc-Sr327}. 
If the system gains an anisotropy of in-plane lattice constants by $x$\%, 
the anisotropy of $t$ is expected around $3.5 x$\% when hybridization between 
the Ru $d$ and O $p$ states is a major contribution to $t$ \cite{harrison}.  
Since a required anisotropy of $t$ to reach a QCEP can become very small [see \fig{fRGresults}(d)], 
not only the rapid drop of the CEL, but also 
the QCEP can be observed in experiments. 

{\it Quasi-one-dimensional metals.} 
A piece of the broken Griffiths wings is also realized for a strong anisotropy 
in Figs.~\ref{fRGresults}(a) and (c).   Such a strong anisotropy is intrinsically 
realized in quasi-one-dimensional metals. In this case, 
when the system is located 
close to van Hove filling, our theory predicts that the anisotropy of the electronic 
system can change dramatically 
by controlling a uniaxial pressure or carrier density. 
Although we are not aware of experiments discussing such 
a phenomenon, there are theoretical works 
reporting it in a different context \cite{quintanilla0910} and their meta-nematic transition 
can be interpreted as originating from a Griffiths wing. 
While $xy$ anisotropy of physical quantities in an already strongly anisotropic system 
was not likely recognized as something related to nematicity, 
we have revealed that the Griffiths wing interpolates between a two- and one-dimensional 
system. Moreover, various microscopic interactions 
can generate an attractive interaction of the $d$PI \cite{yamase00,metzner00,valenzuela01,quintanilla08,lin10}. 
Therefore we may reasonably wait for further experiments. 

{\it Cold atom systems.} 
Each condensed matter system is characterized by a certain 
value of $\mu_{d}$, which is determined by an intrinsic property of the material 
such as a crystal structure and cannot be changed much externally. 
However, $\mu_d$ is fully tunable  for 
optical lattices in a cold atom system \cite{bloch08,quintanilla0910} 
by changing the strength of laser beams between the $x$ and $y$ direction. 
One may employ cold fermions with a large dipolar moment and align the dipole along the $z$ direction. 
Dipolar interaction then yields a repulsive interaction 
between fermions, which leads to an attractive interaction of the $d$PI, 
i.e., the second term in \eq{model}  \cite{valenzuela01,clin10}.  
The Griffiths wings are then expected near van Hove filling 
at temperatures well below the Fermi energy [\fig{fRGresults}(a)]. 
Such low temperatures are now accessible in experiments \cite{aikawa14}.

{\it Anomalous ground state.} 
Our obtained results (\fig{fRGresults}) contain interesting insights into 
electronic nematicity and will likely promote further theoretical studies. 
A non-Fermi liquid ground state is stabilized at 
a quantum critical point of the $d$PI \cite{oganesyan01,metzner03,garst10}. 
It is plausible to expect an anomalous ground state 
also at a QCEP of the $d$PI. 
In an intermediate region between the QCEP$_2$ and QCEP$_{1'}$, 
the dashed line  in \fig{fRGresults}(a), 
quantum fluctuations 
completely wash out the wing even close to van Hove filling. 
If the system remains a Fermi liquid there, 
the density of states would diverge at van Hove filling. 
We would then expect a Griffiths wing there 
because our Hamiltonian (\ref{model}) 
has an attractive interaction of the $d$PI. 
However, we have obtained a crossover around the dashed line in \fig{fRGresults}(a). 
This consideration hints a possible non-Fermi liquid ground state 
at van Hove filling; the same conclusion was obtained also by Dzyaloshinskii 
in a different context \cite{dzyaloshinskii96}. 
Therefore $xy$ anisotropy can lead to an anomalous ground state in 
a wide parameter space spanned by $\mu$ and $\mu_d$, even though 
the quantum phase transition is a first order at zero anisotropy. 
This possibility is very interesting because usually a non-Fermi liquid can be 
stabilized only at a certain point at zero temperature such 
as a quantum critical point, except for purely one-dimensional systems. 

\section{conclusions}
Electronic nematic order couples directly to $xy$ anisotropy. 
The anisotropy can be controlled by applying a uniaxial pressure and a strain. 
Moreover, actual materials often contain intrinsic anisotropy due to a crystal structure such as 
orthorhombicity. The present theory addresses such a situation in a rather general setup 
by including the $d$-wave chemical potential $\mu_{d}$ 
in a low-energy effective Hamiltonian of an electronic nematic  instability [see \eq{model}]. 
Although the nematic physics is frequently discussed in a two-dimensional 
system, the present theory shows that the nematic physics is important 
also in an anisotropic system. 
We have shown that the Griffiths wing associateds with the nematicity 
interpolates between two- and (nearly) one-dimensional systems by 
changing the anisotropy (Figs.~\ref{MFresults} and \ref{fRGresults}). 
In fact, a recent theoretical work \cite{quintanilla0910} reports 
a meta-nematic phenomena in an anisotropic system, which can be 
interpreted as coming from the Griffiths wing. 
The Griffiths wing of the nematicity is found to be quite unique in the sense 
that it exhibits a non-monotonic temperature dependence (\fig{MFresults}), 
in sharp contrast to the cases of He$^{3}$-He$^{4}$ mixtures \cite{griffiths70} 
and ferromagnetic systems \cite{belitz05,kotegawa11,aoki11}.  
The Griffiths wing turns out to be very sensitive to nematic order-parameter 
fluctuations, leading to a phase diagram (\fig{fRGresults}) topologically 
different from the mean-field phase diagram:   
a QCEP close to zero anisotropy, a crossover region, 
and a broken Griffiths wing terminated with two QCEPs in a strong anisotropy. 
This suggests that even if fluctuations are relatively weak 
and the phase diagram is still well captured by mean-field theory at zero anisotropy, 
fluctuation effects can be dramatic once $xy$ anisotropy is introduced. 
Hence the effect of fluctuations 
is definitely important to the Griffiths wing. 
Given that order-parameter fluctuations are present to a greater or lesser extent in 
actual materials, our \fig{fRGresults} is expected to be more realistic than \fig{MFresults}. 
Not only at three QCEPs, but also in the crossover region, the ground state 
may feature non-Fermi liquid behavior. 
This possibility is very interesting since a non-Fermi liquid state 
can extend in a wide parameter space at zero temperature. 
We hope that the present theory serves as a fundamental basis of the nematic physics in 
various materials such as high-$T_c$ cuprates, double-layered ruthenates, quasi-one-dimensional 
metals, and cold atoms, and will promote further theoretical studies.

\acknowledgments
The author appreciates very much insightful and valuable discussions with 
K. Aikawa, F. Benitez, A. Eberlein, T. Enss, 
A. Greco, 
N. Hasselmann, P. Jakubczyk, 
A. Katanin,  W. Metzner, 
M. Nakamura, 
B. Obert, and S. Takei. 
Support by the Alexander von Humboldt Foundation 
and a Grant-in-Aid for Scientific Research from Monkasho 
is also greatly appreciated.

\appendix
%\setcounter{section}{1}
%\section{flow euqaitons}
%\section*{Appendix}
\section{}
%technical details about the analysis of order-parameter fluctuations}
%\section{Supplementary Material}
%\setcounter{page}{1}
In this Appendix we present 
technical details of our fRG framework. 

Our formalism of the fRG partly overlaps with previous works \cite{jakubczyk09,yamase11a},  
which studied how nematic fluctuations change 
the mean-field phase diagram for $\mu_d=0$. 
Compared with previous calculations \cite{jakubczyk09, yamase11a},  
we did the following extension. 
i) Introduction of two cutoffs, one is the physical cutoff $\Lambda_{0}$ 
which gives the upper cutoff to the summation in 
Eq.~(\ref{bareaction}) 
as emphasized by adding the prime 
to the symbol $\Sigma$,  
and the other is the ultraviolet cutoff $\Lambda^{\rm UV}$ which is in principle 
infinite. 
In the previous formalism \cite{jakubczyk09,yamase11a} $\Lambda^{UV}$ 
was assumed to be identical to $\Lambda_{0}$. 
ii) No additional approximations to compute the right hand side of 
flow equations, that is, 
we take account of quantum fluctuations in the anomalous 
dimension $\eta=-\frac{{\partial}\log Z}{{\partial} \log \Lambda}$, 
the contribution from the term $\frac{\partial Z}{\partial \Lambda}$, and 
an additional term coming from 
the momentum derivative of the regular for the flow of $Z$. 

The resulting flow equations become different 
from those in Refs.~\onlinecite{jakubczyk09} and \onlinecite{yamase11a}: 
\bea
&&\hspace{-10mm}
\Lambda \partial_{\Lambda} U(\phi) = \frac{T}{4\pi} \frac{1}{1+\frac{U''}{Z \Lambda^{2}}} 
\left\{
\frac{2-\eta}{2}
  \left[
  \hat{\Lambda}^{2} + 2\sum_{n=1}^{n_{\rm max}}(\hat{q}_{2}^{2} - \hat{q}_{1}^{2}) 
  \right]
  +\frac{\eta}{4\Lambda^2}
  \left[
  \hat{\Lambda}^{4} + 2\sum_{n=1}^{n_{\rm max}}(\hat{q}_{2}^{4} - \hat{q}_{1}^{4}) 
  \right]  
\right\} , \label{flowU}\\
&&\hspace{-10mm} 
\Lambda\partial_{\Lambda} Z = -\eta Z , \label{flowZ}
\eea
and 
\be
\eta= \left\{ 
\begin{array}{l}  \frac{\displaystyle \frac{T}{4\pi} \frac{
(U''')^{2}}{Z^{3}\Lambda^{6}(1+\frac{U''}{Z \Lambda^{2}})^{4}} 
\left[
  {\Lambda}^{2} + 3\displaystyle \sum_{n=1}^{n_{\rm max}} (\hat{q}_{2}^{2} - \hat{q}_{1}^{2})
  \right] 
  }
  {\displaystyle 1+ \frac{T}{8\pi}\frac{
  (U''')^{2}}{Z^{3}\Lambda^{6}(1+\frac{U''}{Z \Lambda^{2}})^{4}} 
\displaystyle \sum_{n=1}^{n_{\rm max}} (\hat{q}_{2}^{2} - \hat{q}_{1}^{2})
\left[ 4-\frac{3}{\Lambda^{2}} (\hat{q}_{2}^{2} + \hat{q}_{1}^{2}) \right]
  }  \quad {\rm for} \, \Lambda \leq \Lambda_{0} , \\
  0 \quad {\rm for} \,  \Lambda > \Lambda_{0} .
  \end{array}
  \right.
\ee
$U'' (U''')$ denotes the second (third) derivative with respect to $\phi$; 
$n_{\rm max}$ is the maximum Matsubara frequency contributing 
to the flow equations and is given by 
$n_{\rm max}=\left[\frac{Z\hat{\Lambda}^{3}}{3\pi\sqrt{3}A_{0}T}\right]$; 
$\hat{q}_{1} (> 0)$ and $\hat{q}_{2} (\geq \hat{q}_{1})$ are real roots of the equation 
\be
q^{3}-\hat{\Lambda}^{2}q +\frac{A_{0}}{Z} |\omega_{n}|=0\, ;
\ee
and 
\be
\hat{\Lambda} = \left\{
\begin{array}{l}
\Lambda \quad  {\rm for} \quad \Lambda \leq \Lambda_{0} , \\
\Lambda_{0} \quad  {\rm for} \quad  \Lambda > \Lambda_{0} .
\end{array}
\right.
\ee

We solve the differential equations (\ref{flowU}) and (\ref{flowZ}) numerically 
by reducing $\Lambda$ from $\Lambda^{\rm UV}$ to zero; the initial condition 
is given by the bare action Eq.~(\ref{bareaction}). 
Since we cannot set $\Lambda^{\rm UV}=\infty$ numerically, 
we first did calculations for various choices of large $\Lambda^{\rm UV}$ and 
checked that our conclusions do not depend on 
the value of $\Lambda^{\rm UV} (\geq \Lambda_{0})$. 
In addition, our conclusions also do not depend on 
a precise choice of $A_{0}$ and $Z_{0}$. 
We took $\Lambda^{\rm UV}=\Lambda_{0}$, $A_{0}=1$, 
and $Z_{0}=10$ in \fig{fRGresults}. 

%%%%%%%%%%%%%%%%%%%%%%%%%%%%%%%%%%%%%%%%%%%%%%%
%%
%%                           REFERENCES
%%
%%%%%%%%%%%%%%%%%%%%%%%%%%%%%%%%%%%%%%%%%%%%%%%

\bibliography{main.bib}% Produces the bibliography via BibTeX.

\begin{thebibliography}{67}
\bibitem[{mis({\natexlab{a}})}]{misc-fig1}
\bibinfo{note}{Figure~\ref{MFresults}(a) does not depend on the value of
  $g(>0)$ and we took $g=1$ in Figs.~\ref{MFresults}(b) and (c).}
\bibitem[{mis({\natexlab{b}})}]{misc-BI}
\bibinfo{note}{In the presence of a small $t'$ ($|t'/t|<0.5$), the BI is
  realized for $\mu=4(t-t')$ for $\mu_{d}<2$ and $\mu=2\mu_{d}+4t'$ for
  $\mu_{d}>2$.}
\bibitem[{mis({\natexlab{c}})}]{misc-largemud}
\bibinfo{note}{If $t'$ is included, the system always allows a hopping along
  the diagonal direction. Hence the anisotropy of $\mu_d \approx 2$ does not
  necessary mean that the system becomes fully one dimensional.}
\bibitem[{mis({\natexlab{d}})}]{misc-Sr327}
\bibinfo{note}{While an $xy$ anisotropy is also generated by introducing a
  magnetic field along the $x$ or $y$ direction \cite{mackenzie12}, the
  coupling to electrons is different from our anisotropic field of $\mu_d$.}
\end{thebibliography}

\end{document}